%% file: kdd.tex
\documentclass[sigconf]{acmart}

\usepackage{pifont}
\usepackage{pbox}
\usepackage{enumitem}

\newcommand{\cmark}{\ding{51}}%
\newcommand{\xmark}{\ding{55}}%

\AtBeginDocument{%
  \providecommand\BibTeX{{%
    \normalfont B\kern-0.5em{\scshape i\kern-0.25em b}\kern-0.8em\TeX}}}

\copyrightyear{2023}
\acmYear{2023}
\setcopyright{rightsretained}
\acmConference[KDD '23]{Proceedings of the 29th ACM SIGKDD Conference on Knowledge Discovery and Data Mining}{August 6--10, 2023}{Long Beach, CA, USA}
\acmBooktitle{Proceedings of the 29th ACM SIGKDD Conference on Knowledge Discovery and Data Mining (KDD '23), August 6--10, 2023, Long Beach, CA, USA}
\acmDOI{10.1145/3580305.3599825}
\acmISBN{979-8-4007-0103-0/23/08}

\makeatletter
\gdef\@copyrightpermission{
    \begin{minipage}{0.3\columnwidth}
    
    \href{https://creativecommons.org/licenses/by-sa/4.0/}{\includegraphics[width=0.90\textwidth]{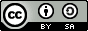}}
    
    \end{minipage}\hfill
    \begin{minipage}{0.7\columnwidth}
    
    \href{https://creativecommons.org/licenses/by-sa/4.0/}{This work is licensed under a Creative Commons Attribution-ShareAlike International 4.0 License.}
    
    \end{minipage}
    \vspace{5pt}
}
\makeatother

\begin{document}

\title{FedMultimodal: \\ A Benchmark For Multimodal Federated Learning}


\author{Tiantian Feng}
\affiliation{%
  \institution{University of Southern California}
  \city{Los Angeles}
  \state{CA}
  \country{USA}
  }
  \email{tiantianf@usc.edu}

\author{Digbalay Bose}
\affiliation{%
  \institution{University of Southern California}
  \city{Los Angeles}
  \state{CA}
  \country{USA}
}
  \email{dbose@usc.edu}

\author{Tuo Zhang}
\affiliation{%
  \institution{University of Southern California}
  \city{Los Angeles}
  \state{CA}
  \country{USA}
}
  \email{tuozhang@usc.edu}

\author{Rajat Hebbar}
\affiliation{%
  \institution{University of Southern California}
  \city{Los Angeles}
  \state{CA}
  \country{USA}
}
  \email{rajatheb@usc.edu}

\author{Anil Ramakrishna}
\affiliation{%
  \institution{Amazon Alexa AI}
  \city{Los Angeles}
  \state{CA}
  \country{USA}
}
  \email{aniramak@amazon.com}

\author{Rahul Gupta}
\affiliation{%
  \institution{Amazon Alexa AI}
  \city{Boston}
  \state{MA}
  \country{USA}
}
  \email{gupra@amazon.com}

\author{Mi Zhang}
\affiliation{%
  \institution{The Ohio State University}
  \city{Columbus}
  \state{OH}
  \country{USA}
}
  \email{mizhang.1@osu.edu}

\author{Salman Avestimehr}
\affiliation{%
  \institution{University of Southern California}
  \city{Los Angeles}
  \state{CA}
  \country{USA}
}
  \email{avestime@usc.edu}

\author{Shrikanth Narayanan}
\affiliation{%
  \institution{University of Southern California}
  \city{Los Angeles}
  \state{CA}
  \country{USA}
}
  \email{shri@sipi.usc.edu}

\renewcommand{\shortauthors}{Tiantian Feng et al.}

\begin{abstract}
Over the past few years, Federated Learning (FL) has become an emerging machine learning technique to tackle data privacy challenges through collaborative training. In the Federated Learning algorithm, the clients submit a locally trained model, and the server aggregates these parameters until convergence. Despite significant efforts that have been made to FL in fields like computer vision, audio, and natural language processing, the FL applications utilizing multimodal data streams remain largely unexplored. It is known that multimodal learning has broad real-world applications in emotion recognition, healthcare, multimedia, and social media, while user privacy persists as a critical concern. Specifically, there are no existing FL benchmarks targeting multimodal applications or related tasks. In order to facilitate the research in multimodal FL, we introduce FedMultimodal, the first FL benchmark for multimodal learning covering five representative multimodal applications from ten commonly used datasets with a total of eight unique modalities. FedMultimodal offers a systematic FL pipeline, enabling end-to-end modeling framework ranging from data partition and feature extraction to FL benchmark algorithms and model evaluation. Unlike existing FL benchmarks, FedMultimodal provides a standardized approach to assess the robustness of FL against three common data corruptions in real-life multimodal applications: missing modalities, missing labels, and erroneous labels. We hope that FedMultimodal can accelerate numerous future research directions, including designing multimodal FL algorithms toward extreme data heterogeneity, robustness multimodal FL, and efficient multimodal FL. The datasets and benchmark results can be accessed at: \href{https://github.com/usc-sail/fed-multimodal}{https://github.com/usc-sail/fed-multimodal}.
\end{abstract}

\ccsdesc[500]{Computing methodologies~Distributed computing methodologies}

\keywords{Federated Learning, Multimodal Learning, Multimodal Benchmark}



\maketitle

\input{sections/intro.tex}

\input{sections/data.tex}
\input{sections/pipeline.tex}

\input{sections/robust.tex}
\input{sections/evaluation.tex}

\bibliographystyle{ACM-Reference-Format}
\bibliography{ref}


\end{document}

%% file: sections/intro.tex
\section{Introduction}


With rapid advances in machine learning (ML) \cite{lecun2015deep} in the past decade, modern mobile devices and wearable sensors \cite{booth2019multimodal, pantelopoulos2009survey} have revolutionized applications and services in industries ranging from entertainment and transportation to healthcare and defense, significantly changing how people live, work, and interact with each other. These intelligent sensing devices, equipped with sensors of multiple modalities, can capture diverse information about a user, including but not limited to physiological, emotional, and rich spatiotemporal contextual information \cite{feng2019discovering, booth2019toward, feng2021multimodal, patel2012review, BecerikGerber2022TheFO}. These data records are typically transmitted to remote servers for centralized training of the ML models. However, the collection of human-centered data raises significant concerns about compromising user privacy due to association with sensitive environments and contexts, ranging from homes, workplaces, and business meetings to hospitals and schools \cite{raij2011privacy}. Therefore, it is critical to ensure that modern ML systems can protect user privacy by preventing any unauthorized access to data \cite{mireshghallah2020privacy, feng2023trust}.

In response to this, ML practitioners have developed Federated Learning as an alternative paradigm to build models, without the need to transfer user data from the edge devices~\cite{konevcny2016federated}. Unlike centralized training, models are trained locally using locally stored data, and updated parameters are transmitted to the server instead of raw data. FL allows clients to train a model collaboratively without sharing their local data, making it one of the most emerging privacy-enhancing learning algorithms in ML research \cite{kairouz2021advances}.

\begin{table}[t]
\caption{Experimental considerations between FedMultimodal and existing multimodal FL studies.}
\vspace{-2mm}
    \small
    \begin{tabular*}{0.9\linewidth}{lccc}
        \toprule
        & \textbf{Missing} & \textbf{Erroneous} & \textbf{Missing}  \\ 
        & \textbf{Modailties} & \textbf{Labels} & \textbf{Labels}  \\ 
        \midrule 
        SSCL \cite{saeed2020federated} & \xmark & \xmark & \cmark \\
        MMFed \cite{xiong2022unified} & \xmark & \xmark & \xmark \\
        FedMsplit~\cite{chen2022fedmsplit} & \cmark & \xmark & \xmark \\
        CreamFL~\cite{yu2023multimodal} & \cmark & \xmark & \xmark \\
        \midrule 
        \textbf{FedMultimodal (Ours)} & \cmark & \cmark & \cmark \\
        \bottomrule
    \end{tabular*}
\vspace{-3mm}
\label{table:comparison_mm}
\end{table}

Previous works in FL have primarily focused on designing robust and efficient algorithms for federated model training. \textit{FedAvg} \cite{mcmahan2017communication} was the earliest FL optimization algorithm to train the model in the distribution mechanism. 
In \textit{FedAvg}, each client executes local model updates before submitting the updates to the server. 
Even though \textit{FedAvg} offers possibilities for deploying FL in the wild, it often encounters slow convergence as a consequence of gradient drifting from data heterogeneity. As such, researchers have proposed algorithms such as stochastic Controlled Averaging Algorithm (SCAFFOLD) \cite{karimireddy2020scaffold} and FedProx \cite{li2020federated} to minimize the impact of gradient drift for heterogeneous data. For example, SCAFFOLD accelerates the training speed through control variates which prevent the client gradients from drifting away from the global optima. Similarly, \cite{reddi2020adaptive} introduced adaptive optimization algorithms, FedOpt, that allow server optimization through momentum. 




To facilitate FL research in more diverse problem domains, a number of FL benchmarks have been developed in the past few years.
For example, LEAF \cite{caldas2018leaf} was the earliest FL benchmark which includes multiple FL training tasks on 5 datasets covering various computer vision and NLP tasks. 
FedML~\cite{chaoyanghe2020fedml}, besides providing an open-source library and a platform for federated learning deployment, it includes multiple FL benchmarks on computer vision and health~\cite{he2021fedcv}, data mining~\cite{He2021FedGraphNNAF}, IoT~\cite{Zhang2021FederatedLF}, and NLP \cite{lin2021fednlp}.
More recently, \cite{lai2021fedscale} announced a multi-domain FL benchmark called FedScale. FedScale included implementations with 20 realistic FL datasets mainly in computer vision and natural language processing applications. 
\cite{dimitriadis2022flute} introduced an FL simulation tool named FLUTE, which covers the application of CV, NLP, and audio tasks.
Meanwhile, \cite{zhang2022fedaudio} presented an audio-centric federated learning framework, FedAudio, which focused on speech emotion recognition, keyword spotting, and audio event classification. Further, FLamby \cite{terrail2022flamby} is a recently proposed FL benchmark for a wide range of healthcare applications such as identifying lung nodules and predicting death risks. FederatedScope~\cite{Xie2022FederatedScopeAC} also incorporates various benchmarks for federated learning in CV, NLP, and data mining~\cite{Wang2022FederatedScopeGNNTA}.

\vspace{1mm}

\noindent \textbf{Existing Multimodal Federated Learning Works:} While existing FL benchmarks largely focus on unimodal applications such as computer vision (CV), natural language processing (NLP), and speech recognition, a significant number of real-world applications are associated with multimodal data streams. 
As listed in Table~\ref{table:comparison_mm}, \cite{saeed2020federated} was one of the earliest to investigate FL using multi-sensory data. They proposed a self-supervised learning approach called Scalogram-signal Correspondence Learning (SSCL) to learn robust multi-modal representations in FL. More recently, \cite{xiong2022unified} designed a multi-modal FL framework named MMFed using the cross-attention mechanism. Moreover, \cite{chen2022fedmsplit} proposed an FL framework called FedMSplit that targeted the issue of missing modalities in the multimodal setup. CreamFL~\cite{yu2023multimodal} provides a multi-modal FL framework using contrastive representation-level ensemble to learn a larger server model from heterogeneous clients across multi-modalities. However, existing multimodal FL frameworks perform their evaluation using their defined experimental setups, thus making it challenging for researchers to compare their methods with existing state-of-the-art fairly and effectively.

\begin{table*}[t]
\caption{Overview of the 10 datasets included in FedMultimodal.}
\vspace{-3mm}
    \small
    \begin{tabular*}{0.97\linewidth}{ccccccccc}
    
        \toprule
       \textbf{Task} & \parbox{2cm}{\textbf{Dataset}} & \textbf{Partition} & \textbf{Client Num.} & \textbf{Modalities} & \textbf{Features} & \textbf{Metirc} & 
        \parbox{1cm}{\centering \textbf{Validation\\Protocol}} & 
        \parbox{1.25cm}{\centering \textbf{Total\\Instance}} \\ 
        \midrule 

        \parbox{1.25cm}{\centering \textbf{ER}} & 
        \parbox{2cm}{MELD \\ CREMA-D } & 
        \parbox{1.5cm}{\centering Natural \\ Natural } & 
        \parbox{0.5cm}{\centering 86 \\ 72 } & 
        \pbox{1.7cm}{Audio, Text \\ Audio, Video} & 
        \pbox{3cm}{MFCCs, MobileBert \\ MFCCs, MobileNetV2} & 
        \parbox{0.75cm}{\centering UAR} & 
        \parbox{1.5cm}{\centering Pre-defined \\ 5-Fold} & 
        \parbox{1.25cm}{\centering 9,718 \\ 4,798} \\
        
        \midrule 
        \parbox{1.25cm}{\centering \textbf{MAR}} & 
        \parbox{2cm}{UCF101 \\ MiT10 \\ MiT51} & 
        \parbox{1.5cm}{\centering Synthetic \\ Synthetic \\ Synthetic } & 
        \parbox{0.5cm}{\centering 100 \\ 200 \\ 2000 } &
        \pbox{1.7cm}{Audio, Video \\ Audio, Video \\ Audio, Video} & 
        \pbox{3cm}{MFCCs, MobileNetV2 \\ MFCCs, MobileNetV2 \\ MFCCs, MobileNetV2} & 
        \parbox{0.75cm}{\centering Top1 \\ Acc} & 
        \parbox{1.5cm}{\centering Pre-defined} & 
        \parbox{1.25cm}{\centering 6,837 \\ 41.6K \\ 157.6K} \\

        \midrule 
        \parbox{1.25cm}{\centering \textbf{HAR}} & 
        \parbox{2cm}{UCI-HAR \\ KU-HAR} & 
        \parbox{1.5cm}{\centering Synthetic \\ Natural} & 
        \parbox{0.5cm}{\centering 105 \\ 66} &
        \pbox{1.7cm}{Acc, Gyro \\ Acc, Gyro} & 
        \pbox{3cm}{Raw \\ Raw} & 
        \parbox{0.75cm}{\centering F1} & 
        \parbox{1.5cm}{\centering Pre-defined \\ 5-Fold} & 
        \parbox{1.25cm}{\centering 8,979 \\ 10.3K} \\

        \midrule 
        \parbox{1.25cm}{\centering \textbf{Health}} & 
        \parbox{2cm}{PTB-XL} & 
        \parbox{1.5cm}{\centering Natural} & 
        \parbox{0.5cm}{\centering 34 } &
        \pbox{1.8cm}{I-AVF, V1-V6} & 
        \pbox{3cm}{Raw} & 
        \parbox{0.75cm}{\centering F1} & 
        \parbox{1.5cm}{\centering Pre-defined} & 
        \parbox{1.25cm}{\centering 21.7K} \\

        \midrule 
        \parbox{1.25cm}{\centering \textbf{SM}} & 
        \parbox{2cm}{Hateful-Memes \\ CrisisMMD} & 
        \parbox{1.5cm}{\centering Synthetic \\ Synthetic} & 
        \parbox{0.5cm}{\centering 50 \\ 100 } &
        \pbox{1.8cm}{Image, Text} & 
        \pbox{3cm}{MobileNetV2, MobileBert \\ MobileNetV2, MobileBert} & 
        \parbox{0.75cm}{\centering AUC \\ F1} & 
        \parbox{1.5cm}{\centering Pre-defined \\ Pre-defined} & 
        \parbox{1.25cm}{\centering 10.0K \\ 18.1K} \\
        
        \bottomrule
    \end{tabular*}
\vspace{-0mm}
\label{table:dataset}
\end{table*}

\vspace{1mm}
\noindent
\textbf{Our Contributions:} In this work, we introduce FedMultimodal, a FL benchmark for multimodal applications. We summarize our key contributions as follows: 

\begin{itemize}[leftmargin=*]
\item FedMultimodal includes \textbf{ten representative datasets} covering \textbf{five diverse application scenarios} – emotion recognition, multimodal action recognition, human activity recognition, healthcare, and social media – that are well aligned with FL. We present systematic benchmark results on the above datasets to facilitate researchers to fairly compare their algorithms.
\vspace{1mm}
\item To help the community accurately compare performance and ensure \textbf{reproducibility}, FedMultimodal presents an \textbf{open-source end-to-end FL simulation framework} and includes capabilities to perform data partitioning, feature processing, and multimodal training. 
FedMultimodal offers support for several popular FL optimizers including FedAvg \cite{mcmahan2017communication}, FedProx \cite{li2020federated}, FedRS \cite{li2021fedrs}, SCAFFOLD \cite{karimireddy2020scaffold}, and FedOpt \cite{reddi2020adaptive}, and provide flexibility that allows users to customized the trainers on the included datasets. The source codes and user guides are available at \textbf{\url{ https://github.com/usc-sail/fed-multimodal}}.
\vspace{1mm}

\item In addition to ensuring accessibility and reproducibility, the benchmark provides a \textbf{robustness assessment module} that allows researchers to simulate challenges uniquely tied to multimodal FL applications in real-world scenarios. As listed in Table~\ref{table:comparison_mm}, previous works on multimodal FL provide limited assessments of the robustness under real-world settings. Specifically, FedMultimodal emulates \textbf{missing modalities}, \textbf{missing labels}, and \textbf{erroneous labels} on top of the provided datasets to simulate scenarios when deploying FL systems in real-world settings. This is a crucial difference and a unique contribution of FedMultimodal compared to existing FL literature.
\end{itemize}




%% file: sections/data.tex
\section{Multimodal Datasets and Tasks}
\label{sec:dataset}
Table~\ref{table:dataset} provides an overview of the 10 datasets included in
FedMultimodal. These 10 multimodal datasets cover five diverse tasks -- Emotion Recognition, Multimedia Action Recognition, Human Activity Recognition, Healthcare, and Social Media classification. One important reason we select these 10 datasets is that they are publicly available, thus ensuring ease of accessibility and reproducibility. In this section, we provide a brief overview of each included dataset and the corresponding tasks.


\subsection{Emotion Recognition (ER)}
Emotion recognition (ER) has broad applicability of ER in virtual assistant-based tasks, human behavior analysis, and AI-assisted education, making it a valuable research topic in FL \cite{tsouvalas2022privacy,feng2022semi}. FedMultimodal benchmark incorporates two widely used datasets in this category: MELD and CREMA-D.

\vspace{1mm}
\noindent \textbf{MELD} is a multiparty dialog dataset \cite{poria2018meld} containing over 9k utterances with audio and transcripts data from the Friends TV series. Due to the imbalanced label distribution in the dataset, we keep 4 emotions with the most samples i.e., neutral, happy, sad, and angry.

\vspace{1mm}
\noindent \textbf{CREMA-D} has 7,442 audio-visual clips recorded by 91 actors \cite{cao2014crema}. Each speaker was instructed to utter 12 sentences emulating 6 emotions: neutral, happy, anger, disgust, fear, and sad.

\subsection{Multimodal Action Recognition (MAR)} 
The task of MAR consists of classifying a video into action categories based on underlying visual and audio modalities. In FedMultimodal, we include two well-known MAR testbeds: UCF101 and Moments in Time (MiT).


\vspace{1mm}
\noindent \textbf{UCF101} dataset \cite{soomro2012ucf101} consists of 13,320 web videos with 101 sport-based action labels. However, data associated with only 51 labels are presented with video and audio information, resulting in less than 7,000 videos for the experiments. The duration of the videos ranges from several seconds to over 20 seconds. We subsample the video at the frame rate of 1Hz to reduce the computation overhead.

\vspace{1mm}
\noindent \textbf{Moments in Time (MiT)} is a large-scale MAR (~1 million) dataset \cite{monfort2019moments} with short (3 seconds) videos with overall list of 339 action labels. It is worth noting that MiT is a challenging dataset, with state-of-the-art top-1 accuracy close to 35\% \cite{ryoo2019assemblenet}. Given the inherent difficulty of this task, we tackle the easier classification problem by creating partitions of data with fewer distinct labels. We create two sub-datasets, MiT10 and MiT51, from the original MiT dataset. MiT10 and MiT51 contain videos of the 10 and the 51 most frequent labels. Similar to the UCF101 setting, we subsample the video every 10 frames to accommodate the computing constraints in FL.
\subsection{Human Activity Recognition (HAR)}
HAR identifies human actions based on wearable data such as accelerometers and gyroscopes. Due to the nature of its wearable-friendly attribute, it has become a prevalent research topic in FL \cite{sannara2021federated}. FedMultimodal provides the implementation on two HAR datasets: UCI-HAR and KU-HAR. In our experiments, we treat the accelerometer and gyroscope data as two different modalities.

\vspace{1mm}
\noindent \textbf{UCI-HAR} dataset \cite{anguita2013public} consists of smartphone sensors (Accelerometer and Gyroscope) data from 30 subjects (19-48 yrs old) performing six daily activities: walking, walking upstairs, walking downstairs, sitting, standing, laying. The participants wear smartphones on their waists during the collection phase. The accelerometer and gyroscope data are sampled at 50Hz.

\vspace{1mm}
\noindent \textbf{KU-HAR} is a recent human activity recognition dataset \cite{sikder2021ku} collected with a group of 90 participants (75 male and 15 female) on 18 different activities. Instead of evaluating the 18 activities, we decided to keep 6 activities existing in the UCI-HAR dataset while adding jumping and running activities. 

\begin{figure*}
    \centering
    \includegraphics[width=0.95\linewidth]{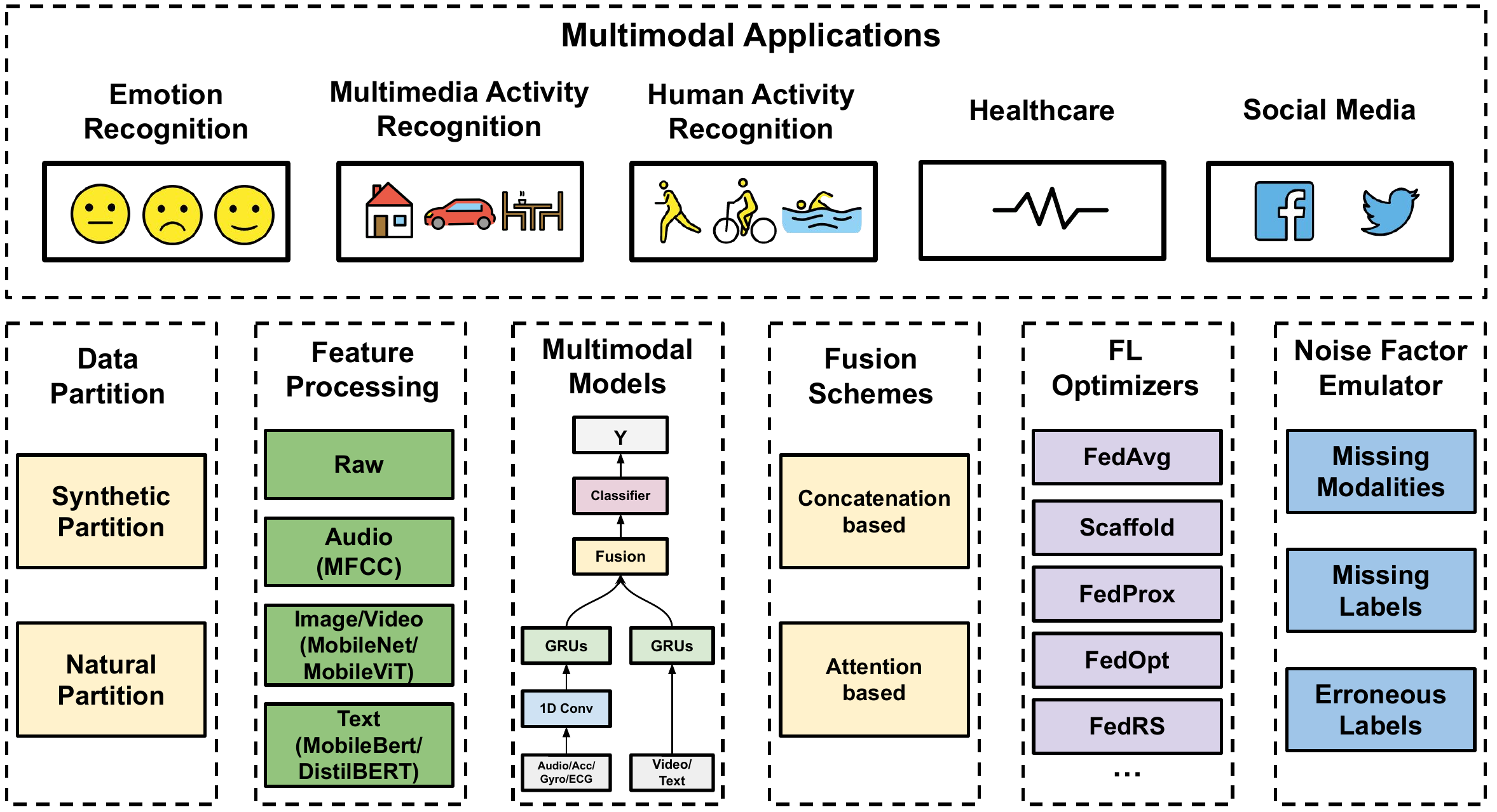}
    \caption{The overall architecture of the end-to-end multimodal federated learning framework included in FedMultimodal.}
    \label{fig:fed_multimodal}
    \vspace{-1mm}
\end{figure*}

\subsection{Healthcare}
Healthcare ML applications have made immense progress in a range of domains e.g., heart-disease classification over the last decade. FedMultimodal explores the problem of ECG classification based on the PTB-XL \cite{wagner2020ptb} dataset. 
\textbf{PTB-XL} \cite{wagner2020ptb} includes over 20,000 clinical 12-lead ECG recordings from 18,885 patients for a multi-label classification task. There are 5 classes describing ECG diagnosis, including normal ECG, myocardial infarction, ST/T change, conduction disturbance, and hypertrophy. As suggested by \cite{strodthoff2020deep}, we use the ECG data provided at the sampling frequency of 100 Hz. We separate the readings from electrodes I, II, III, AVL, AVR, AVF, and V1-V6 as two modalities suggested by \cite{alday2020classification}. 
\subsection{Social Media (SM)}
Social media has become an increasingly important tool during disasters and emergencies for people to track the latest updates in the area, especially the impact (e.g., property damage, injury, and death) of the disaster, as well as urgent needs for help. However, the widespread adoption of social media has also drawn significant concerns about spreading misinformation, thus urging the need to identify and mitigate this misleading and harmful content. To accelerate FL research in this domain, FedMultimodal incorporates two social-media-based multimodal datasets related to hateful content and crisis information classification. 

\vspace{1mm}
\noindent \textbf{Hateful Memes} dataset was released from the 2020 Hateful Memes Challenge \cite{kiela2020hateful} that focus on detecting hateful speech in memes. The database includes 10,000 multimodal data with image and text pairs with binary classes. 

\vspace{1mm}
\noindent \textbf{CrisisMMD} \cite{alam2018crisismmd} comprises 18.1k tweets containing both paired visual and textual information. It collects relevant tweets from seven prominent natural disasters such as California Wildfires (2017). One of the purposes of the dataset is to identify the impact of the disaster like utility damage and injured or dead people.

%% file: sections/pipeline.tex
\section{End-to-end Multimodal Federated Learning Framework}
To benchmark the performance of the multimodal datasets described in Section~\ref{sec:dataset} as well as to support future research in the area of multimodal federated learning, we have built an end-to-end multimodal federated learning research framework.
Figure~\ref{fig:fed_multimodal} \footnote{Figure~\ref{fig:fed_multimodal} uses image sources from https://openmoji.org/} illustrates the overall architecture of the framework. 
As shown, our framework covers the complete pipeline of multimodal federated learning, which includes six key components: (1) non-IID data partitioning, (2) feature processing, (3) multimodal models, (4) fusion schemes, (5) federated optimizers, and (6) real-world noise factor emulator. 
In particular, one key difference between FedMultimodal and existing multimodal FL literature is that FedMultimodal takes the real-world noise factors into consideration and examines model robustness to three real-world noise factors: missing modalities, missing labels, and erroneous labels. 
%
In this section, we describe each of the six key components in detail.

\subsection{Non-IID Data Partitioning}
The non-IID data partitioning is a fundamental step in emulating FL experiments. The first partition scheme is through the unique client identifier. For example, speech-related datasets, like CREMA-D and MELD, comprise speech-text or speech-visual data organized by speaker IDs. Hence, it is natural to use speaker IDs to partition the client data in FL, creating authentic non-IID data distributions. Similarly, we consider partitions in datasets like KU-HAR and PTB-XL comprise data with based on participant IDs and clinical site IDs, respectively.
On the other hand, other multimodal datasets used in this paper, including MAR and SM datasets, do not have such realistic client partitions thus requiring ML practitioners to synthesize non-IID data distributions. Following prior works, we partition these datasets using Dirichlet distribution with $\alpha\in\{0.1, 5.0\}$ to control the level of data heterogeneity, where $\alpha=0.1$ and $\alpha=5.0$ represents high heterogeneity and low heterogeneity, respectively. Although the original UCI-HAR datasets consist of data partitioned by participants, each participant performed the same amount of activities from each category, making the label distribution IID in UCI-HAR. Hence, we increase the heterogeneity of data distribution by dividing each participant's data using the Dirichlet distribution. 

\subsection{Feature Processing}
Instead of training the model using raw input data like images and texts from scratch, FedMultimodal leverages well-established pre-trained models as backbone networks to extract features for the training downstream models. Unlike the centralized training paradigm, feature processing in federated learning benchmarks demands considerations in computation efficiency and feasibility. Mainly, the selected feature needs to align with the computation capabilities available on the edge computing devices. For example, it is unrealistic to assume that edge devices could load and run large transformer-based \cite{transformers} models for inference purposes without sacrificing system performance. Hence, we focus on implementing mobile-friendly feature extraction pipelines in FedMultimodal which are listed below, targeting swift computation, efficient storage, and ease of deployment.

\begin{itemize}[leftmargin=*]
    \item \textbf{Visual}: For the visual data, our benchmark supports \textbf{MobileNetV2} \cite{howard2017mobilenets} and \textbf{MobileViT} \cite{mehta2021mobilevit} as the embedding network to extract latent presentations. The complete MobileNetV2 and MobileViT have 4.3M and 2.7M parameters, respectively, making them practical visual feature backbones in FL. Due to space constraints, we report benchmark results with MobileNetV2 in this paper.
    \item \textbf{Text}: FedMultimodal integrates both \textbf{MobileBERT} \cite{sun2020mobilebert} and \textbf{DistillBERT} \cite{sanh2019distilbert} to extract representations from textual data. MobileBERT uses a bottleneck structure to reduce the parameter size from 340M to 25M when compared to BERT \cite{Devlin2019BERTPO}, while DistillBERT applies a knowledge distillation process that decreases the BERT model to 66M parameters. We decide to benchmark with the MobileBERT feature backbone given the page constraints.
    \item \textbf{Audio}: FedMultimodal uses Mel-frequency cepstral coefficients (MFCCs) due to their widespread usage in the state-of-the-art speech recognition models like Wav2Vec 2.0 \cite{chen2021exploring}.
    \item \textbf{Other Modalities}: We use the raw data with the remaining modalities in the FedMultimodal. These data streams include accelerometer, gyroscope, and ECG readings.
\end{itemize}

\begin{figure}
    \centering
    \includegraphics[width=0.75\linewidth]{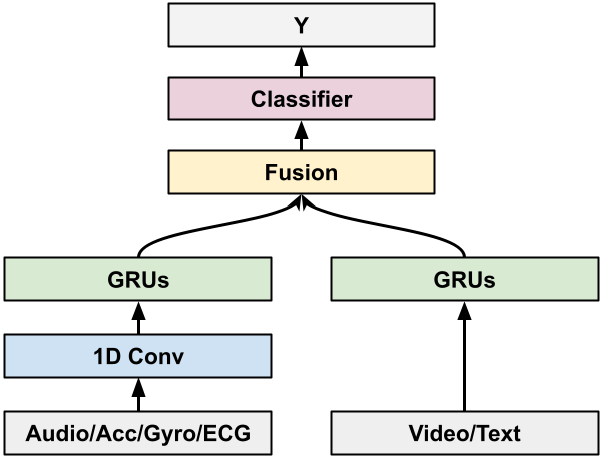}
    \caption{The architecture of the basic model.}
    \label{fig:multimodal_model}
    \vspace{-3mm}
\end{figure}
\input{sections/Multimodal.tex}

%% file: sections/Multimodal.tex
\subsection{Multimodal Models}

\vspace{1mm}
\noindent \textbf{Model Design Principles.} 
Compared to remote servers, edge computing nodes are more appropriate for lightweight computing tasks due to constraints in computation resources, storage capabilities, battery capacities, and communication bandwidths. When designing ML models for resource-constrained devices, a significant design consideration is to reduce the number of parameters in edge ML models, thus reducing memory and execution latency. Such models can either be the backbone feature extraction models or application-specific prediction models. One major design principle of FedMultimodal is to study lightweight but effective solutions for multimodal FL learning instead of training models with multi-million parameters. 


\vspace{1mm}
\noindent \textbf{Model Architecture.} 
With this design principle in mind, we construct ML models mainly based on the 1D Conv-RNN/MLP architecture. Even though the transformer-based model has achieved SOTA performance in diverse applications, these models typically include millions or even billions of parameters, making them impractical to use in FL settings as a result of massive computations, memory usage, and battery consumption during back-propagation \cite{wang2020survey}.

An example model architecture is presented in Figure~\ref{fig:multimodal_model}. Specifically, the multimodal model in FedMultimodal includes an encoder, a modality-fusion block, and a downstream classifier. The encoder part follows either Conv+RNN architecture or RNN-only architecture. The encoder that adopts Conv+RNN architecture takes the input of audio, accelerometer, gyroscope, and ECG information, otherwise uses RNN-only architecture. Following encoder modules, FedMultimodal uses a late-fusion mechanism to combine modality-specific representations into a multimodal representation. The multimodal representation is then fed through 2 dense layers for downstream predictions.

\begin{table}[t]
    \caption{Hyperparameters for training FL models with multimodal data using FedAvg.}
    \small
    \begin{tabular*}{0.97\linewidth}{cccc}
    
        \toprule
        
        \parbox{1.9cm}{\textbf{Multimodal\\ Dataset}} & 
        \parbox{2cm}{\centering \textbf{Client\\Sample Rate}} & 
        \parbox{1.5cm}{\centering \textbf{Learning\\Rate}} & 
        \parbox{2cm}{\centering \textbf{Training Rounds}} \\ 

        \midrule

        \parbox{1.9cm}{MELD \\ CREMA-D } & 
        \parbox{2cm}{\centering 10\% \\ 10\% } & 
        \parbox{1.25cm}{\centering 0.01 \\ 0.05 } & 
        \parbox{1.7cm}{\centering 200 \\ 200} \\

        \parbox{1.9cm}{UCF101 \\ MiT10 \\ MiT51} & 
        \parbox{2cm}{\centering 10\% \\ 5\% \\ 5\% } & 
        \parbox{1.25cm}{\centering 0.05 \\ 0.05 \\ 0.05 } &
        \parbox{1.7cm}{\centering 200 \\ 300 \\ 300} \rule{0pt}{5ex} \\

        \parbox{1.9cm}{UCI-HAR \\ KU-HAR} & 
        \parbox{2cm}{\centering 10\% \\ 10\%} & 
        \parbox{1.25cm}{\centering 0.05 \\ 0.05} &
        \parbox{1.7cm}{\centering 200\\200} \rule{0pt}{3.5ex} \\

        \parbox{1.9cm}{PTB-XL} & 
        \parbox{2cm}{\centering 25\%} & 
        \parbox{1.25cm}{\centering 0.05 } &
        \parbox{1.7cm}{\centering 200} \rule{0pt}{2.5ex} \\
        
        \parbox{1.9cm}{Hateful-Memes \\ CrisisMMD} & 
        \parbox{2cm}{\centering 25\% \\ 10\%} & 
        \parbox{1.25cm}{\centering 0.05 \\ 0.05 } &
        \parbox{1.8cm}{\centering 200 \\ 200} \rule{0pt}{3.25ex} \\
        
        \bottomrule
    \end{tabular*}
\vspace{-2mm}
\label{table:experiment_setup}
\end{table}

\begin{table*}[t]
\caption{Benchmarking performance. Text colors in red and blue denote the best performance using \textit{Concatanation-based Fusion} and \textit{Attention-based Fusion}, respectively. $\dagger$ indicates the best performance score of the corresponding dataset.}
    \small
    \begin{tabular*}{0.97\linewidth}{lccccccccccc}
    
        \toprule
        & & & &
        \multicolumn{4}{c}{\textbf{Concatanation-based Fusion}} & 
        \multicolumn{4}{c}{\textbf{Attention-based Fusion}} \\ 
        \cmidrule(lr){5-8} \cmidrule(lr){9-12}
        
        & \parbox{2cm}{\textbf{ML Datasets}} & 
        $\mathbf{\alpha}$ & 
        \textbf{Metric} & 
        \textbf{FedAvg} & 
        \textbf{FedProx} & 
        \textbf{FedRS} & 
        \textbf{FedOpt} & 
        \textbf{FedAvg} &
        \textbf{FedProx} & 
        \textbf{FedRS} & 
        \textbf{FedOpt} \\ 
        \midrule 

        \parbox{1.75cm}{\centering \textbf{Natural\\Partition}} & 
        \parbox{2cm}{MELD \\ CREMA-D \\ KU-HAR \\ PTB-XL} & 
        - &
        \parbox{1.0cm}{\centering UAR \\ UAR \\ F1 \\ F1} &
        \parbox{1.0cm}{\centering  48.08 \\ 61.52 \\ 61.56 \\ 61.87 } & 
        \parbox{1.0cm}{\centering  48.47 \\ 61.64 \\ 61.13 \\ 61.00} & 
        \parbox{1.0cm}{\centering  49.21 \\ \textbf{\color{red}62.17} \\ 61.26 \\ -} & 
        \parbox{1.0cm}{\centering  \textbf{\color{red}50.66} \\ 61.14 \\ \textbf{\color{red}68.82} \\ \textbf{\color{red}62.83$^\dagger$} } & 
        \parbox{1.0cm}{\centering 54.37 \\ 61.66 \\ 61.78 \\ 61.88 } & 
        \parbox{1.0cm}{\centering 54.67 \\ 62.03 \\ 61.78 \\ 61.71 } & 
        \parbox{1.0cm}{\centering 53.82 \\ 60.41 \\ 62.04 \\ - } & 
        \parbox{1.0cm}{\centering \textbf{\color{blue}55.37$^\dagger$} \\ \textbf{\color{blue}62.66$^\dagger$} \\ \textbf{\color{blue}71.41$^\dagger$} \\ \textbf{\color{blue}62.42} } \\

        \midrule
        \parbox{1.75cm}{\centering \textbf{Synthetic\\Partition}} & 
        \parbox{2cm}{UCF101 \\ MiT10 \\ MiT51 \\ Hateful-Memes \\ CrisisMMD \\ UCI-HAR} & 
        5.0 & 
        \parbox{1.0cm}{\centering Acc \\ Acc \\ Acc \\ AUC \\ F1 \\ F1} &
        \parbox{1.0cm}{\centering  67.98 \\ 55.39 \\ 28.96 \\ 58.23 \\ 43.67 \\ 78.78 } & 
        \parbox{1.0cm}{\centering  67.98 \\ 55.39 \\ 28.62 \\ \textbf{\color{red}59.90$^\dagger$} \\ 43.37 \\ 78.07 } & 
        \parbox{1.0cm}{\centering  67.98 \\ 55.29 \\ 27.67 \\ 57.98 \\ \textbf{\color{red}44.30$^\dagger$} \\ 77.87 } & 
        \parbox{1.0cm}{\centering  \textbf{\color{red}74.38} \\ \textbf{\color{red}55.47} \\ \textbf{\color{red}35.01} \\ 58.97 \\ 43.44 \\ \textbf{\color{red}84.83} } & 
        \parbox{1.0cm}{\centering  75.13 \\ 57.10 \\ 33.90 \\ 57.83 \\ 39.11 \\ 77.75} & 
        \parbox{1.0cm}{\centering  74.51 \\ \textbf{\color{blue} 57.93$^\dagger$} \\ 34.46 \\ 59.09 \\ 39.36 \\ 77.38 } & 
        \parbox{1.0cm}{\centering  75.27 \\ 56.82 \\ 33.74 \\ 56.67 \\ \textbf{\color{blue}41.01} \\ 76.82 } & 
        \parbox{1.0cm}{\centering  \textbf{\color{blue}75.89$^\dagger$} \\ 57.25 \\ \textbf{\color{blue}35.62$^\dagger$} \\ \textbf{\color{blue}59.51} \\ 38.74 \\ \textbf{\color{blue}85.17$^\dagger$}} \\

        \midrule
        \parbox{1.75cm}{\centering \textbf{Synthetic\\Partition}} & 
        \parbox{2cm}{UCF101 \\ MiT10 \\ MiT51 \\ Hateful-Memes \\ CrisisMMD \\ UCI-HAR} & 
        0.1 & 
        \parbox{1.0cm}{\centering Acc \\ Acc \\ Acc \\ AUC \\ F1 \\ F1} &
        \parbox{1.0cm}{\centering  64.57 \\ 44.84 \\ 28.63 \\ 51.02 \\ 9.90 \\ 77.50 } & 
        \parbox{1.0cm}{\centering  64.55 \\ 50.03 \\ 27.98 \\ \textbf{\color{red}59.86} \\ 10.65 \\ 77.34 } & 
        \parbox{1.0cm}{\centering  61.17 \\ 45.92 \\ 27.92 \\ 51.28 \\ 9.28 \\ 73.68 } & 
        \parbox{1.0cm}{\centering  \textbf{\color{red}74.17} \\ \textbf{\color{red}50.10} \\ \textbf{\color{red}33.46} \\ 58.08 \\ \textbf{\color{red}26.82} \\ \textbf{\color{red}78.97}} & 
        \parbox{1.0cm}{\centering  74.53 \\ 42.96 \\ 32.41 \\ 49.68 \\ 8.49 \\ 76.66 } & 
        \parbox{1.0cm}{\centering  74.71 \\ 45.47 \\ 32.55 \\ 59.44 \\ 25.31 \\ 76.58 } & 
        \parbox{1.0cm}{\centering  73.27 \\ 46.23 \\ 31.99 \\ 49.80 \\ 10.12 \\ 68.65 } & 
        \parbox{1.0cm}{\centering  \textbf{\color{blue}75.05$^\dagger$} \\ \textbf{\color{blue}50.76$^\dagger$} \\ \textbf{\color{blue}35.35$^\dagger$} \\ \textbf{\color{blue}60.51$^\dagger$} \\ \textbf{\color{blue}27.59$^\dagger$} \\ \textbf{\color{blue}79.80$^\dagger$}} \\

        \bottomrule
    \end{tabular*}
\vspace{-1mm}
\label{table:baseline_results}
\end{table*}

\subsection{Fusion Schemes}
In this work, we present two basic fusion approaches: \textit{concatenation-based fusion} and \textit{attention-based fusion}. In the concatenation-based fusion, the average pooling operation is first performed on the GRU output. After that, we concatenate the pooling embeddings to form the multimodal embedding. On the other hand, the attention-based fusion concatenates the temporal output from each modality without the average pooling step. We apply an attention mechanism similar to hierarchical attention \cite{yang2016hierarchical}. Given the concatenated multimodal data $h$, the attention procedures are as follows:
\vspace{-1mm}
\begin{align*}
    u = tanh(Wh + &b); a = softmax(u^{T}c) \\
    v &= \sum_{i}a_{i}h_{i}
\end{align*}
\vspace{-1.5mm}

The concatenated multimodal data $h$ is first fed through a one-layer MLP to get representation $u$. We then use a context vector $c$ to obtain a normalized importance score through a softmax function. After that, we compute the final multimodal embedding $v$ as a weighted sum of $h$ based on the weights $a$. Here, we can further implement a multi-head attention mechanism by having multiple $c$. We would also stress that this attention mechanism is lightweight, thus making it realistic to deploy on a variety of edge devices. In addition, the attention mechanism allows us to mask the missing modalities in the computation, providing a simple yet effective solution to train FL models with missing modalities.

\subsection{Federated Optimizers}

First, most existing FL training algorithms are validated in unimodal settings, and their efficacy on multimodal tasks remains unexplored. As a result, FedMultimodal is suited to several popular FL algorithms, including FedAvg \cite{mcmahan2017communication}, FedProx \cite{li2020federated}, FedRS \cite{li2021fedrs}, and FedOpt \cite{reddi2020adaptive}. Particularly, FedOpt holds state-of-the-art performance across multiple unimodal applications \cite{reddi2020adaptive}. One objective of FedMultimodal is to provide comprehensive evaluations across various FL algorithms.

%% file: sections/robust.tex
\subsection{Real-world Noise Factor Emulator}

Prior literature (see Table~\ref{table:comparison_mm}) on multimodal FL provides little or no assessment of their robustness in real-life settings. In order to provide a comprehensive evaluation of multimodal FL models toward safe and robust deployment, FedMultimodal enables the emulation of missing modalities, missing labels, and erroneous labels for real-world multimodal FL.

\vspace{0.75mm}
\noindent \textbf{Missing Modality.} In practice, data sources, whether they are microphones, cameras, mobile hardware, or medical electrodes, are prone to data imperfections or complete data losses (e.g., missing modalities) caused by firmware malfunctions, network disconnections, or sensor damages \cite{feng2019imputing}. Hence, it is critical to design an emulator module to synthesize the cases of missing modalities for some clients. FedMultimodal provides the simulations of missing modalities as suggested by \cite{chen2022fedmsplit}, where the availability of each modality follows a Bernoulli distribution. We set an equal missing rate $q$ for each modality in the following experiments.

\vspace{0.75mm}
\noindent \textbf{Missing Labels.} Not only can the multimodal FL encounter data imperfection challenges, it can also suffer from missing label problems. Surprisingly, most prior works have made the ideal assumption that the data stored on edge devices are fully annotated with ground-truth labels. However, in a more realistic real-world FL setting, we argue that only a portion of the data can come with labels. As such, FedMultimodal allows the missing label simulation to assess the risk of decreased robustness.

\vspace{0.75mm}
\noindent \textbf{Erroneous Labels.} In addition to missing labels, real-world FL implementations encounter label noise as a result of bias, skill differences, and labeling errors from the annotators. Inspired by \cite{zhang2022fedaudio}, we apply a label error generation process described in \cite{northcutt2021confident}. In summary, the erroneous labels are generated using a transition matrix $Q$, where $Q_{i,j} = \mathbb{P}(y=j|y=i)$ indicates the chance of ground-truth label $i$ being erroneous annotated with label $j$. 

%% file: sections/evaluation.tex
\begin{figure*}
    \centering
    \includegraphics[width=0.93\linewidth]{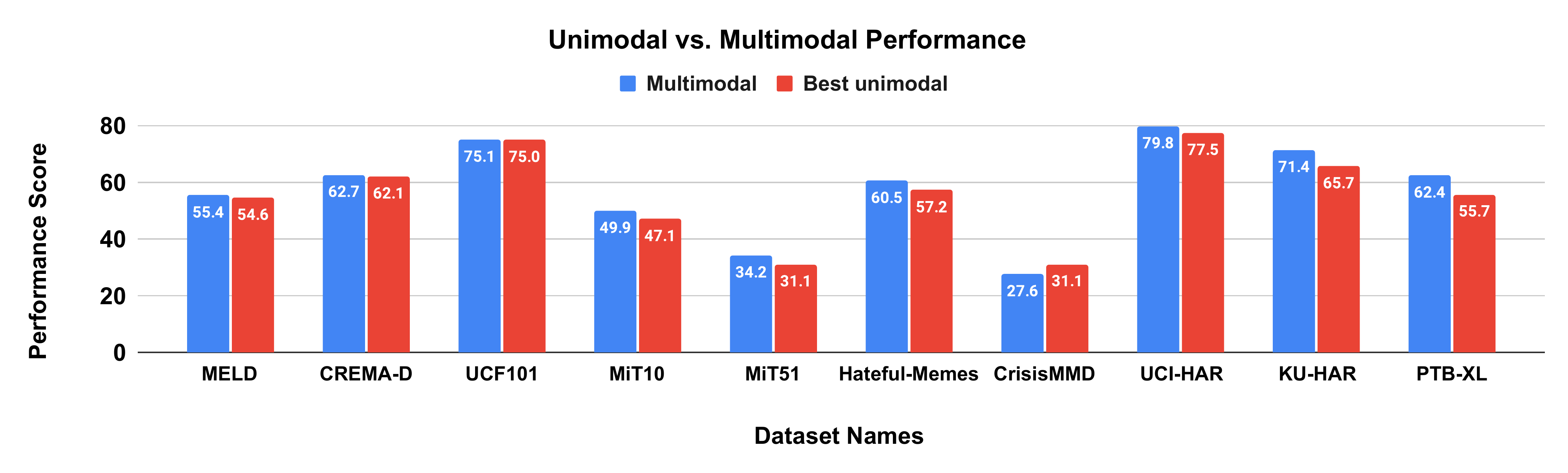}
    \vspace{-4mm}
    \caption{Performance comparisons between multimodal and unimodal learning under FL settings.}
    \label{fig:unimodal_vs_multimodal}
\end{figure*}

\section{Experiments and Discussion}

\subsection{Experimental Details}

\noindent
\textbf{Setup.}
We adopt the RNN-only model architecture to the video and text modalities while utilizing the Conv-RNN model architecture in other modalities. Specifically, the model with the convolutional module consists of 3 convolution layers with the number of filters in \{16, 32, 64\} and the filter kernel size of $5 \times 5$. Moreover, we set the hidden layer size of RNN as 128. We choose ReLU as the activation function and the dropout rate as 0.2. The number of attention heads is 6 in all experiments. We fixed the batch size for all datasets to 16 and the local epoch to 1 for all experiments.

Additionally, we set the training epochs as 200 for all datasets except the MiT sub-datasets. However, the total training epoch is 300 in MiT10 and MiT51 as these 2 datasets contain more data than the other datasets. Hyperparameter details such as the learning and client sampling rates for each dataset are listed in Table~\ref{table:experiment_setup}. Due to the limited number of clients in the Hateful Memes dataset and PTB-xl datasets, we apply a higher client sample rate in these 2 datasets. In the FedOpt algorithm, we search the server learning rate in a range from $10^{-3}$ to $2.5 \times 10^{-3}$. Meanwhile, the proximal term ranges from $10^{-2}$ to $10^{0}$ in the FedProx algorithm.
\begin{figure*}
    \centering
    \includegraphics[width=0.95\linewidth]{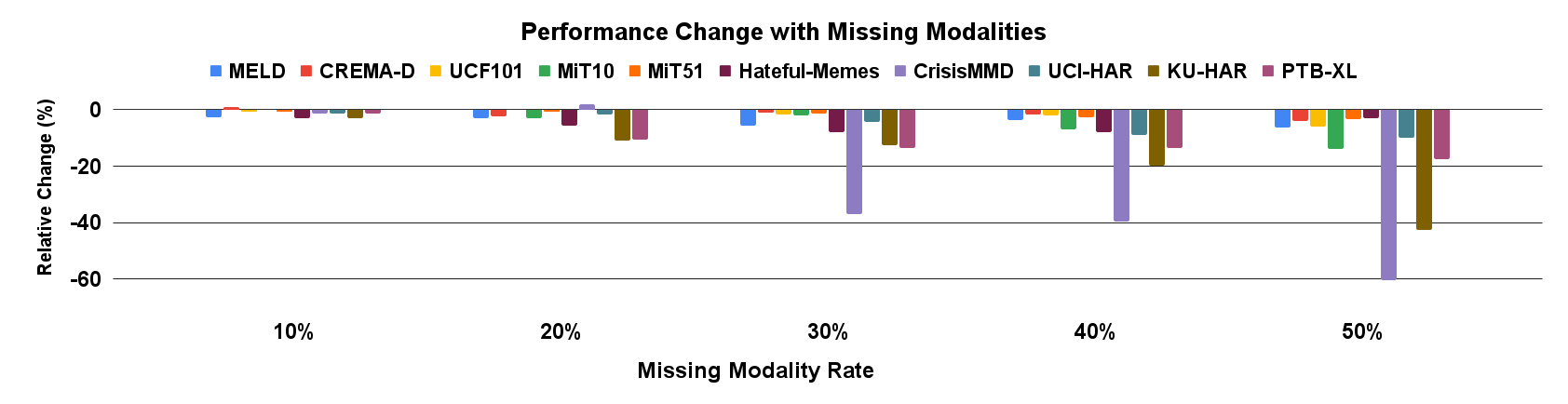}
    \vspace{-4mm}
    \caption{Relative performance changes under different missing modality rates.}
    \label{fig:missing_modality}
\end{figure*}

\vspace{1mm}
\noindent
\textbf{Evaluation Metrics.}
We follow established practices from the literature while conducting evaluations on each dataset. Specifically, evaluation metrics (e.g., F1) and validation protocols (e.g., predefined splits) are two fundamental components to ensure comparability with past (and future) works. With datasets that provide a pre-defined partition for training/validation/testing, we repeat the experiments 5 times using different seeds. We perform 5-fold cross-validation on datasets without such pre-defined experimenting rules. We provide details about our evaluation methods in Table~\ref{table:dataset}.

\begin{figure*}
    \centering
    \includegraphics[width=0.92\linewidth]{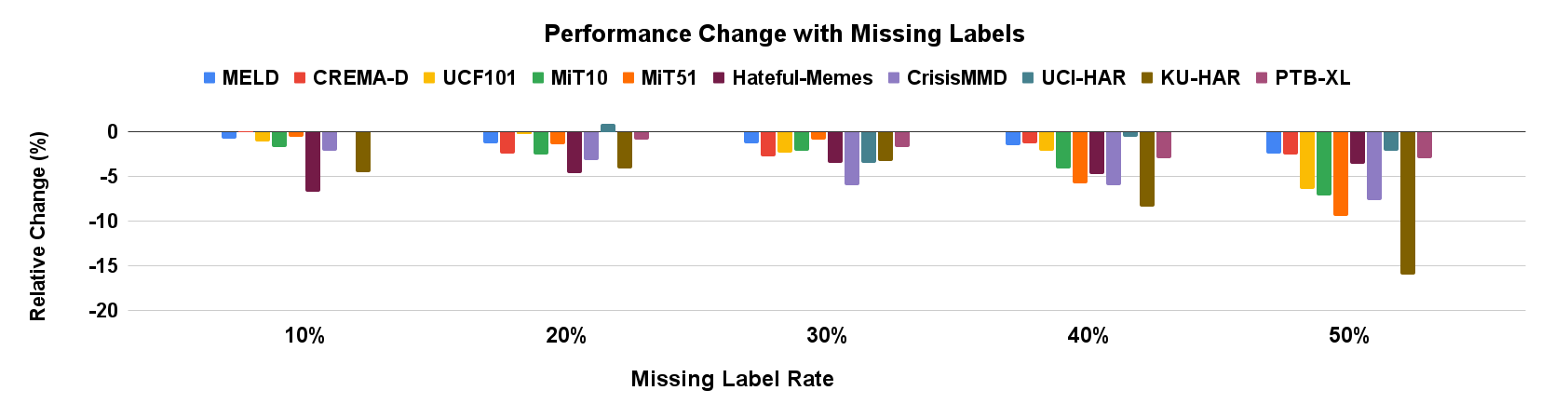}
    \vspace{-2mm}
    \caption{Relative performance changes under different label missing rates.}
    \label{fig:missing_labels}
    \vspace{-2mm}
    
\end{figure*}

\begin{figure*}
    \centering
    \includegraphics[width=0.9\linewidth]{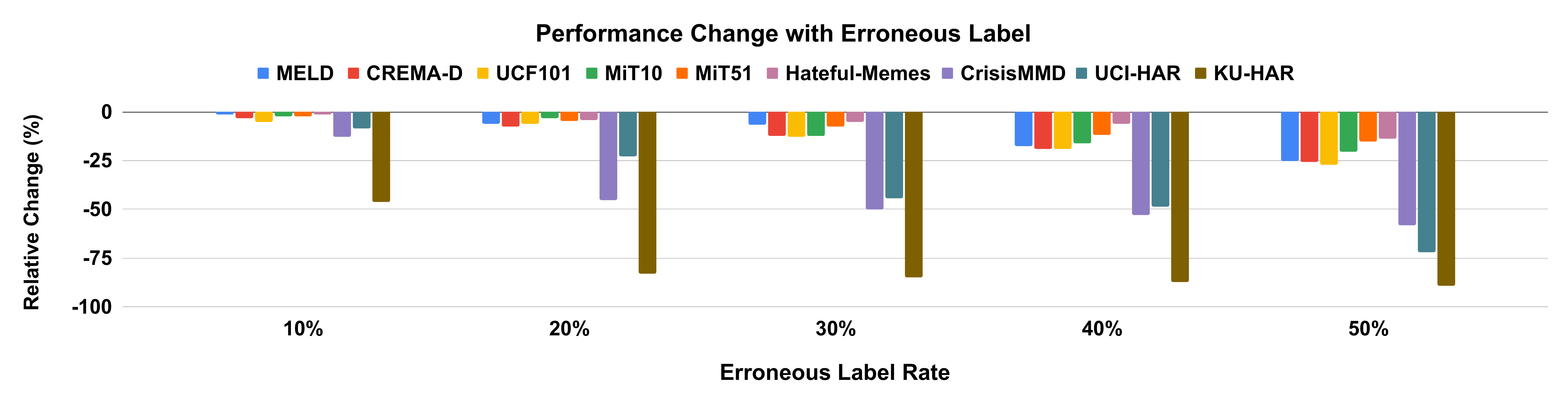}
    \vspace{-2mm}
    \caption{Relative performance changes under different erroneous label rates.}
    \label{fig:noisy_labels}
\end{figure*}

\subsection{Overall Performance}
%
We first report the comparisons between two fusion mechanisms (\textbf{attention-based fusion} and \textbf{concatenation-based fusion}) in Table~\ref{table:baseline_results}. From the results, we can find that the attention-based fusion mechanism outperforms concatenation-based fusion in the majority of the datasets. Specifically, the attention-based fusion mechanism leads to better performances in most high data heterogeneity conditions, but it underperforms the concatenation-based fusion in two synthetic datasets with $\alpha=5.0$ (low data heterogeneity). Moreover, we observe that the FedOpt algorithm consistently yields better baselines compared to other FL algorithms with a few exceptions in low data heterogeneity conditions. However, we would like to highlight that, in practice, FedOpt requires additional hyperparameter tuning on the server learning rate to reach the best performance. Overall, these results imply that the fusion mechanism is a critical factor impacting multimodal model performance in data heterogeneous FL.

Moreover, HAR tasks are associated with the highest performance scores, suggesting the simplicity of this learning task. In contrast, classification results on the Hateful Memes dataset and CrisisMMD dataset imply that social media classification is a challenging task using FL, with the best model performance on the CrisisMMD dataset below 30\%. A plausible explanation is that the pre-trained models that we rely on are not generalized to social media data, generating image and textual features that are unrepresentative of downstream learning models. On the other hand, performances on the MiT51 dataset demonstrate similar findings pointed out from \cite{monfort2019moments}, validating that MiT is a challenging dataset. However, reducing the number of labels indeed simplifies the predicting task, resulting in moderate model performance on MiT10.

\vspace{-0mm}
\subsection{Uni-modality vs. Multi-modalities}

One fundamental research question centering around multimodal learning is its performance compared to unimodal models. For example, the previous multimodal benchmark \cite{liang2021multibench}, with an emphasis on centralized learning setup, demonstrates that unimodal learning could yield similar performance to multimodal models with fewer parameters. Similar to MultiBench, FedMultimodal provides the unimodal FL to compare with multimodal FL baselines. We summarize the benchmark comparisons between unimodal FL and multimodal FL in Figure~\ref{fig:unimodal_vs_multimodal}. The comparisons use datasets with natural partitions or high data heterogeneity partitions. Overall, we observe that unimodal learning provides competitive performance compared with the multimodal FL benchmarks, complying with centralized benchmark results reported in \cite{liang2021multibench}. Nevertheless, in most scenarios, multimodal learning still outperforms unimodal learning, whereas the performance gap between multimodal and unimodal FL is within 5\% in the majority of the datasets.


\subsection{Impact of Missing Modalities}
As described in earlier sections, a unique challenge associated with multimodal learning is dealing with scenarios of missing modalities \cite{yu2023multimodal, chen2022fedmsplit}. In this section, we benchmark our selected datasets with different rates of missing modalities. In this experiment, we assume that the availability of each modality follows a Bernoulli distribution with a missing rate of $q$. Following the experiment protocol presented by \cite{chen2022fedmsplit}, we set a uniform missing rate of $q$ for each modality, where $q\in\{0.1, 0.2, 0.3, 0.4, 0.5\}$. As described in the multimodal model section, attention-based fusion allows us to train the model even with missing data through masking. To train the model with the missing entries, we fill the missing data with 0 \cite{parthasarathy2020training} while masking out the corresponding data points in calculating attention scores.

\begin{figure*}
    \centering
    \includegraphics[width=0.92\linewidth]{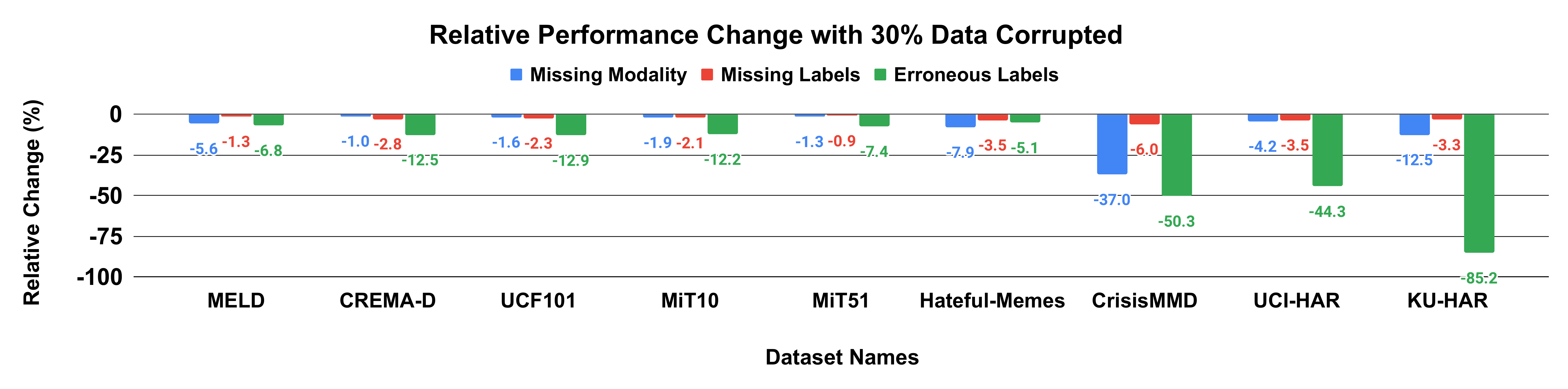}
    \vspace{-5mm}
    \caption{Relative performance changes with 30\% data corrupted (missing modalities vs. missing labels vs. erroneous labels).}
    \label{fig:mm_vs_ml_vs_nl}
    \vspace{-1mm}
\end{figure*}

We present the relative model performance changes at different missing modality rates in Figure~\ref{fig:missing_modality}. From the graph, we find that the relative performance changes with missing rates below 30\% are not substantial, suggesting that a small amount of missing modalities in deployment does not impact the final model performance in multimodal FL. Furthermore, we observe that the model performance starts to decline substantially at the missing rate of 50\%. Surprisingly, we observe that half of the models have relative performance decreases that are under 10\%, suggesting that the provided baseline models that use attention-based fusion still learn useful information in these cases. However, we find that the missing modality introduces a significantly larger impact on CrisisMMD data and the HAR applications compared to the other multimodal applications we evaluated. This suggests that future FL HAR research should carefully consider missing modalities as a part of the evaluation.


\subsection{Impact of Missing Labels}

Missing labels is a widely presented challenge in FL. In this section, we perform evaluations of missing label conditions using the FedMultimodal benchmark. Similar to the missing modality experiment, we assume that the availability of each label follows a Bernoulli distribution with a missing rate $l$. We apply the FedMultimodal benchmark to emulate the missing label rate $l\in\{0.1, 0.2, 0.3, 0.4, 0.5\}$. The goal of this experiment is to quantify the impact of missing labels on the overall performance of benchmarks, hence we do not integrate any mitigation methodologies, such as semi-supervised learning or self-supervised learning in our experiments.

Results on relative model performance changes at different label missing ratios using the FedMultimodal framework are presented in Figure~\ref{fig:missing_labels}. Overall, we can observe that missing labels have a reduced impact on the model performance when compared to the missing modality scenario. For instance, the model performance suffers less than 10\% relative performance decreases in the majority of the datasets with the exception of KU-HAR datasets. When the missing label ratio is below 50\%, we observe minor performance decreases in all datasets. Surprisingly, we identify that CrisisMMD yields worse performance at 30\% than at 10\% missing label ratio. We conjecture that the reason behind this result might be attributed to the data labeling quality of the Hateful-Memes dataset. For example, determining whether the content is hateful or not can be very subjective, and such subjectiveness could hurt labeling quality.

\subsection{Impact of Erroneous Labels}

Besides missing modalities and missing labels, erroneous labels frequently exist in FL~\cite{Yaldiz2023SecureFL}. In this section, we report our benchmark performance with erroneous labels. Similar to previous experiments, we search the erroneous label ratio $l\in\{0.1, 0.2, 0.3, 0.4, 0.5\}$, where $l$ represents the amount of data with erroneous labels. Similar to the experiment setup in \cite{zhang2022fedaudio}, our benchmark defines the sparsity of erroneous label transition matrix $\mathbf{Q}$ as 0.4. The error sparsity specifies the possible number of unique labels $k$ that one label can be wrongly annotated with, with a small sparsity error rate corresponding to a larger $k$. 


The complete results of the relative model performance changes at different levels of erroneous label ratios are shown in Figure~\ref{fig:noisy_labels}. Compared to the missing modalities experiment, the erroneous label condition leads to substantially larger performance decreases. For example, more than half of the datasets have the relative performance decreases above 10\% at the erroneous label ratio of 30\%. Moreover, a 20\% performance drop can be identified from these datasets when the erroneous label ratio reaches 50\%. To compare the impact of data corruption conditions in FL, we plot the relative performance changes with different data corruption conditions at the data corrupted ratio of 30\% in Figure~\ref{fig:mm_vs_ml_vs_nl}. We can observe that performances of multimodal FL are more susceptible to label noises than missing modalities or missing labels. Based on these observations, our future benchmark directions also include implementations of backdoor attacks and mitigation in FedMultimodal.

\section{Limitations and Future Work}

\vspace{0.5mm}
\noindent {\textbf{Scale of Datasets and Models.} The dataset selection criteria of FedMultimodal ensures that the chosen datasets are representative and diversified across different dimensions such as application scenarios, data size, and number of clients. In addition, FedMultimodal only includes ML models that align with the use cases of FL, taking into account the computational limitations of edge devices. 
We acknowledge that FedMultimodal currently does not cover several promising multimodal applications, such as medical image analysis, autonomous driving, and virtual reality, and the range of the supported models is limited. We will continuously update FedMultimodal to support new tasks such as Ego4D \cite{grauman2022ego4d}, as well as newer feature extraction models.}

\vspace{0.5mm}
\noindent {\textbf{Scale of Modality Fusion Schemes.} Currently, FedMultimodal includes two basic approaches for modality fusion: concatenation and attention. Modality fusion under FL remains an open problem, and our objective is to draw attention to the need for developing more advanced modality fusion schemes under FL \cite{lu2016hierarchical, jaegle2021perceiver}.}

\vspace{0.5mm}
\noindent

\noindent \textbf{Data Heterogeneity.} As discussed in previous sections, addressing the data heterogeneity challenge is critical in FL. While many FL studies focus their experiments on the unimodal setup, there is a lack of extensive research on tackling data heterogeneity in multimodal FL. To address this gap, the FedMulitmodal benchmark provides opportunities to facilitate fundamental research in this direction. In the future, it is of further interest to explore knowledge-transfer learning approaches as suggested in \cite{cho2022heterogeneous,lin2020ensemble,itahara2021distillation}, within the context of multimodal FL.

\vspace{0.5mm}
\noindent    
\noindent \textbf{Label Scarcity.} One major challenge for FL is the lack of qualitative labels. FedMultimodal enables researchers to efficiently perform experiments on multimodal FL with missing labels by providing the ability to emulate experimental conditions with missing labels. We hope FedMultimodal brings unique benefits for ML practitioners to develop self-supervised learning \cite{zhuang2021collaborative, zhang2020federated, dong2021federated} and semi-supervised learning \cite{zhang2021improving, kang2022fedcvt, zhao2020semi} algorithms under multimodal FL.

\vspace{0.5mm}
\noindent    
\noindent \textbf{Privacy Leakage.} While sharing model updates is considered to be more private than sharing raw data, recent works have revealed that FL can still be susceptible to privacy attacks. These attacks include (but are not limited to) membership inference attacks \cite{melis2019exploiting}, reconstruction attacks \cite{zhu2020deep, geng2021towards}, attribute inference attacks \cite{feng2021attribute} and label inference attacks \cite{fu2022label}. Consequently, an emerging research direction for expanding FedMultimodal is to explore the privacy leakages in multimodal FL. Apart from identifying privacy risks associated with multimodal FL, it is also crucial to investigate privacy-enhancing techniques, such as differential privacy \cite{wei2020federated, dwork2006differential, feng22b_interspeech} and secure aggregation \cite{bonawitz2017practical} as promising areas of research within the scope of FedMultimodal to mitigate privacy attacks.

\section{Conclusion}

In this paper, we presented a new framework for multimodal federated learning, named FedMultimodal, which enables federated learning in multimodal applications. We further established a reproducible benchmark of results for 5 multimodal FL applications covering 10 datasets for future comparisons. We also benchmarked results on model robustness to missing modalities, missing labels, and noisy labels in each of these tasks.

\section{Acknowledgement}

This work is in part supported by USC Amazon Center, as well as research gift awards from Intel, Meta, and Konica Minolta.